\documentclass[reprint,superscriptaddress,nofootinbib,amsmath,amssymb,prd,floatfix]{revtex4-1}
\pdfoutput=1
\usepackage{graphicx}
\usepackage[dvipsnames]{xcolor}
\usepackage{amsmath}
\allowdisplaybreaks
\usepackage{amssymb}

\usepackage{slashed}
\usepackage[utf8]{inputenc}
\usepackage[colorlinks=true,linkcolor=blue,bookmarksopen,bookmarksnumbered]{hyperref}
\usepackage{color}
\usepackage[normalem]{ulem}
\usepackage{soul}
\usepackage{units}
\usepackage{rotating}
\usepackage{hhline,multirow,tabularx}
\usepackage{bm}
\usepackage{hyperref}
\usepackage[export]{adjustbox}
\usepackage{mdframed}
\usepackage{comment}
\usepackage{natbib}

\def\bea#1\eea{\begin{align}#1\end{align}} 

\newcommand{\bef}{\begin{figure}[htb]\centering}
\newcommand{\eef}{\end{figure}}

\newcommand{\shao}[1]{\marginpar{\footnotesize\textbf{SHAO}}}

\def\<{\langle}
\def\>{\rangle}

\def\cos{\hbox{cos}}
\def\sin{\hbox{sin}}
\def\ln{\hbox{ln}}

\usepackage{lipsum}
\begin{document}

\title{Azimuthal asymmetries of muon pair production in ultraperipheral heavy ion collisions}

\author{Ding Yu Shao}
\email{dingyu.shao@cern.ch}
\affiliation{Department of Physics, Center for Field Theory and Particle Physics, Fudan University, Shanghai, 200433, China}
\affiliation{Key Laboratory of
Nuclear Physics and Ion-beam Application (MOE), Fudan University, Shanghai, 200433, China}

\author{Cheng Zhang}
\email{chengzhang\_phy@fudan.edu.cn}
\affiliation{Department of Physics, Center for Field Theory and Particle Physics, Fudan University, Shanghai, 200433, China}

\author{Jian~Zhou}
\email{jzhou@sdu.edu.cn}
 \affiliation{\normalsize\it Key Laboratory of
Particle Physics and Particle Irradiation (MOE),Institute of
Frontier and Interdisciplinary Science, Shandong University,
QingDao, China }

\author{Ya-jin Zhou}
\email{zhouyj@sdu.edu.cn}
\affiliation{\normalsize\it Key Laboratory of Particle Physics and
Particle Irradiation (MOE),Institute of Frontier and
Interdisciplinary Science, Shandong University, QingDao,  China }


\begin{abstract}
In this paper we study azimuthal asymmetries of the muon pair production in ultraperipheral heavy ion collisions within the joint impact parameter and transverse momentum dependent framework. The final state QED radiation effects are resummed to all orders in perturbation theory, where the complete muon mass corrections are also taken into account. We further make numerical estimations  for azimuthal asymmetries in the different kinematic regions accessible at RHIC and LHC with the derived resummation formula. We find that the lepton mass effects can give sizable corrections to the asymmetries at relatively large pair transverse momentum at RHIC energy.  
\end{abstract}
\maketitle

\section{Introduction}\label{Introduction}
The study of pure electromagnetic(EM) di-lepton production in ultraperipheral heavy ion Collisions(UPCs) has a long history~\cite{Breit:1934zz,CERES:1995vll,STAR:2018ldd,STAR:2018xaj,ATLAS:2018pfw,ALICE:2018ael,Bertulani:1987tz,Bertulani:2005ru,Baltz:2007kq,Klein:2016yzr,ALICE:2022bii,Zhou:2022gbh,Wang:2022ihj,Niu:2022cug}. At low pair transverse momentum,  di-lepton production through coherent photon fusion process in UPCs is enhanced by the factor $Z^2$ where $Z$ is the nuclear charge number.  Due to the  high luminosity and the clean background,  di-lepton production in UPCs offers a unique opportunity to  search for the physics beyond the standard model~\cite{DELPHI:2003nah,ATLAS:2022ryk,Knapen:2016moh,CMS:2018erd,Ellis:2017edi,Xu:2022qme}. It is also proven to be the golden channel to address the novel aspects of QED under extreme conditions ~\cite{Baur:1998ay,Klein:2020fmr,Steinberg:2021lfm,Hattori:2020htm,Copinger:2020nyx,Brandenburg:2021lnj} in high energy scatterings.   More importantly, the  measurements of di-lepton production in UPCs provide a precise calibration necessary for the photons as sources for the photonuclear processes and set a  baseline for the EM probe of the quark-gluon plasma as well.

The photons participating in UPC events are predominately coherent ones with  transverse-momentum $k_\perp \lesssim 1/R$ (30 MeV) where $R$ is the nuclear radius.   The equivalent photon approximation (EPA)~\cite{Bertulani:1987tz,Baltz:2007kq,Bertulani:2005ru,Klein:2016yzr} is commonly applied to describe $k_\perp$ distribution of coherent photons.  
However, to account for the impact parameter (the transverse distance of the two colliding nuclei) dependent transverse momentum distribution of the lepton pair observed in peripheral collisions and UPCs at RHIC and LHC, one has to go beyond this naive EPA method and employ a more sophisticated  formalism~\cite{Vidovic:1992ik,Hencken:1994my} incorporating $b_\perp$ dependence of photon distribution.   Theoretical efforts~\cite{Zha:2018tlq,Klein:2020jom,Wang:2021kxm,Wang:2022gkd,Klusek-Gawenda:2020eja,Wang:2022gkd,Lin:2022flv,Wang:2022ihj,Klusek-Gawenda:2020eja} made along this line turn out to give a rather satisfactory description of the measured $b_\perp$ dependent  mean value of the total transverse momentum of lepton pair. 

On the other hand, the lepton pair can acquire transverse momentum transfer due to the recoil effect caused by the final state  soft photon radiation. Such soft photon contribution to the transverse momentum distribution can be computed in the perturbation theory and has been resummed to all orders up to the leading logarithmic accuracy in Refs.~\cite{Klein:2018fmp,Klein:2020jom}. At low $k_\perp$, the lepton pair transverse momentum distribution is dominated by the primordial coherent photon distribution, while the soft photon contribution yields the perturbative tail at high $k_\perp$. Especially for the muon pair production, the finite mass corrections would be sizeable in the large $k_\perp$ region where the invariant mass of the muon pair is of the same order of the lepton mass. A similar effect was studied in the transverse momentum resummation for heavy quark pairs production \cite{Zhu:2012ts,Li:2013mia,Catani:2014qha,Ju:2022wia}.

It was recently realized that the coherent photons are highly linearly polarized with the polarization vector being parallel to its transverse momentum direction~\cite{Li:2019yzy,Li:2019sin,Xiao:2020ddm,Zhao:2022dac}. A sizable $\cos \, 4\phi$ azimuthal asymmetry in di-electron production induced by  linearly polarized  coherent photons  was observed in a  STAR measurement~\cite{Adam:2019mby}.  A remarkable agreement between the computed asymmetry (16.5\%)~\cite{Li:2019yzy,Li:2019sin} and  the measured asymmetry (16.8\%$\pm$2.5\%) in UPCs has been reached.  With it being experimentally confirmed~\cite{Adam:2019mby,Brandenburg:2022tna}, the linearly polarized photon beam in UPCs provides us a new tool to  explore novel QCD phenomenology~\cite{Hagiwara:2020juc,Xing:2020hwh,Hagiwara:2021qev,Brandenburg:2022jgr,Mantysaari:2022sux,Wu:2022exl}. For example, the linearly polarized photons can give rise to the significant $\cos \, 2\phi$ and   $\cos\,{4\phi}$ modulations in diffractive $\rho^0$ and $J /\psi$ production~\cite{Xing:2020hwh,Zha:2020cst,STAR:2022wfe,Brandenburg:2022jgr}. A recent analysis showed that the distinctive diffractive pattern exhibited in the transverse momentum dependent $\cos \, 2\phi$ asymmetry  is sensitive to the nuclear geometry, the quantum interference effect~\cite{Klein:1999gv,Abelev:2008ew,Xing:2020hwh,Zha:2018jin}, and the production mechanism (coherent/incoherent). Moreover, the $\cos\,{4\phi}$ asymmetry in diffractive $\rho^0$ production in UPCs could give access to the elusive gluon elliptic Wigner distribution~\cite{Hagiwara:2021qev}. 

In this work, we investigate the   $\cos \, 2\phi$ and   $\cos\,{4\phi}$ azimuthal asymmetries in di-muon production, for which case the lepton mass effect can not be neglected.  In particular, the $\cos \, 2\phi $ asymmetry is proportional to lepton mass. Though the observed $\cos \, 2\phi $ asymmetry  in di-electron production at RHIC is consistent with zero at low pair transverse momentum, it is expected to be sizable in di-muon production. In addition to the contribution from the linearly polarized coherent photons, such azimuthal asymmetries also can be generated perturbatively as the final state soft photons are tended to be emitted aligning with the lepton direction.  We  take into account this pure perturbative origin of the asymmetries by employing the resummation established in Refs.~\cite{Zhu:2012ts,Li:2013mia,Catani:2014qha,Catani:2017tuc,Hatta:2020bgy,Hatta:2021jcd}.  We further argue that the azimuthal asymmetry in the large $k_\perp$ region could provide a new opportunity to test our understanding of the resummation formalism beyond the double logarithmic approximation. 

The paper is structured as follows. We derive the azimuthal-dependent di-muon production cross section in the next section. The soft photon contribution is resummed to all orders up to the next to leading logarithmic accuracy. We present the numerical result in section III. The paper is summarized in section IV.

\section{Theoretical setup}\label{theory}
To calculate observables we consider the production of  muon pairs via the photon-photon fusion process in UPCs. We specify the kinematics by writing,
$$
\gamma\left(x_{1} P+k_{1 \perp}\right)+\gamma\left(x_{2} \bar{P}+k_{2 \perp}\right) \rightarrow l^{+}\left(p_{1}\right)+l^{-}\left(p_{2}\right),
$$
where the leptons are produced nearly back-to-back with total transverse momentum $q_{\perp} \equiv p_{1 \perp}+p_{2 \perp}$ being much smaller than $P_{\perp} \equiv (p_{1 \perp}-p_{2 \perp})/2$. In this work, we concentrate on the low $q_\perp$ region where muon pairs are dominantly produced by the coherent photons. As pointed out in Refs.~\cite{Li:2019yzy,Li:2019sin,Xiao:2020ddm}, the polarization vectors of the incoming photons are parallel to their transverse momenta when the longitudinal momentum fractions carried by photons are small. The corresponding photon distributions can be parametrized in terms of the unpolarized photon TMD and     linearly polarized photon TMD in the conventional TMD factorization. However, once we introduce the impact parameter dependence in the cross section calculation which is essential to account for the measured $b_\perp$ dependent behavior of the di-lepton pair transverse momentum, the transverse momentum carried by the incoming photon appears in the amplitude is no longer identical to that in the conjugate amplitude. One then has to go beyond the TMD factorization to accommodate such $b_\perp$ dependence.  Notice that the time like DVCS process also contributes the di-lepton production~\cite{Pire:2008ea}. But in the kinematics under consideration, one can neglect the contribution from this channel. 

Following the formalism developed in Refs.~\cite{Vidovic:1992ik,Hencken:1994my}, we  compute the joint $b_\perp$ and $q_\perp$ dependent di-muon production cross section at the lowest order of QED. The cross section can be cast into the form,
\begin{align}\label{born}
 \frac{d \sigma_{0}}{d^{2} q_{\perp} d^{2} P_{\perp} d y_{1} d y_{2} d^{2} b_{\perp}} = A_0+A_2 \cos 2 \phi +A_4 \cos 4 \phi,
\end{align}
where $\phi$ is the angle between transverse momentum $q_{\perp}$ and $P_{\perp}$.
$y_{1}$ and $y_{2}$ are muon and anti-muon's rapidities, respectively.  These $\cos \, 2\phi$ and $\cos \, 4\phi $ azimuthal modulations are induced by the linearly polarized coherent photons as mentioned earlier.

 The coefficients $A_0, A_2$ and $A_4$ contain the  convolutions of various photon distribution amplitudes. In order to show their expressions in a concise way, we introduce the following shorthand notation,
\begin{eqnarray}
{\cal \int}[d{\cal K}_\perp ]&\equiv& \int d^{2}k_{1\perp}d^{2}k_{2\perp}d^{2}k_{1\perp}'d^{2}k_{2\perp}'e^{i(k_{1\perp}-k_{1\perp}')\cdot b_{\perp}} \nonumber \\&\times& \delta^{2}(k_{1\perp}+k_{2\perp}-q_{\perp}) \delta^{2}(k_{1\perp}'+k_{2\perp}'-q_{\perp}) \\
 & \times& \mathcal{F}(x_1,k_{1\perp}^{2})\mathcal{F}(x_2, k_{2\perp}^{2})\mathcal{F}(x_1, k_{1\perp}'^{2})\mathcal{F}(x_2, k_{2\perp}'^{2}), \nonumber 
\end{eqnarray}
where  $k_{1\perp}$ and $k_{2\perp}$ are the photons' transverse momenta in the amplitude, while  $k_{1\perp}'$ and $k_{2\perp}'$ are the ones in the conjugate amplitude. The longitudinal momentum fractions are fixed according to the external kinematics: $
x_{1} \simeq\sqrt{(P_{\perp}^{2}+m^{2})/s}\left(e^{y_{1}}+e^{y_{2}}\right),
x_{2} \simeq\sqrt{(P_{\perp}^{2}+m^{2})/s}\left(e^{-y_{1}}+e^{-y_{2}}\right)$,  with $s$, $m$ being the center of mass energy and the muon mass, respectively. The function $\mathcal{F}(k_{1\perp}^{2},x_{1})$ describes the probability amplitude for a photon carrying a given momentum. It can be related to the normal photon TMD:  $|\mathcal{F}(k_{1\perp}^{2},x_{1})|^2=x_1 f(x_1,k_{1\perp}^2)$. One notices that the $b_\perp$ dependence enters the cross section via the phase $e^{i(k_{1\perp}-k_{1\perp}')\cdot b_{\perp}}$.

The coefficients $A_0, A_2$ and $A_4$ can then be expressed as,
\begin{widetext}
\begin{eqnarray}
A_{0}\! & = & \!{\cal \int}[{d\cal K}_\perp]\frac{1}{\left(P_{\perp}^{2}+m^{2}\right)^{2}} \Bigl[-2m^{4}\cos\left(\phi_{k_{1\perp}}+\phi_{k_{1\perp}'}-\phi_{k_{2\perp}}-\phi_{k_{2\perp}'}\right)+m^{2}\left(M^{2}-2m^{2}\right)\cos\left(\phi_{k_{1\perp}}-\phi_{k_{1\perp}'}-\phi_{k_{2\perp}}+\phi_{k_{2\perp}'}\right)\nonumber\\
 &  & +P_{\perp}^{2}\left(M^{2}-2P_{\perp}^{2}\right)\cos\left(\phi_{k_{1\perp}}-\phi_{k_{1\perp}'}+\phi_{k_{2\perp}}-\phi_{k_{2\perp}'}\right)\Bigr],  \\
A_{2} \! & = & \!{\cal \int}[{d\cal K}_\perp]\frac{8m^{2}P_{\perp}^{2}}{ \left(P_{\perp}^{2}+m^{2}\right)^{2}} \cos\left(\phi_{k_{1\perp}}-\phi_{k_{2\perp}}\right)\cos\left(\phi_{k_{1\perp}'}+\phi_{k_{2\perp}'}-2\phi\right),\\
A_{4} \! & = & \!{\cal \int}[{d\cal K}_\perp]\frac{-2P_{\perp}^{4}}{\left(P_{\perp}^{2}+m^{2}\right)^{2}}\cos\left(\phi_{k_{1\perp}}+\phi_{k_{1\perp}'}+\phi_{k_{2\perp}}+\phi_{k_{2\perp}'}-4\phi\right).
\end{eqnarray}
\end{widetext}
where $M$ is the invariant mass of the muon pair.  $\phi_{k_{1\perp}}$ is the azimuthal angel between $P_\perp$ and $k_{1\perp}$. Other azimuthal angles are defined in a similar way. As compared to the previous results obtained in Ref.~\cite{Li:2019sin}, we keep the full lepton mass dependence in the hard coefficients in this work.  One sees that the $\cos\, 2\phi$ azimuthal asymmetry is proportional to the lepton mass.  This asymmetry is negligibly small in di-electron production at low $q_\perp$, while it is sizable in di-muon production at RHIC energy, as shown below.  If one carries out the $b_\perp$ integration from 0 to $\infty$, the above results reduce to that computed in TMD factorization~\cite{Li:2019yzy}.

At the tree level, the lepton pair transverse momentum is equal to $q_\perp=k_{1\perp}+k_{2\perp} $ due to momentum conservation.  However, the soft photon radiation effect can significantly modify the lepton pair transverse momentum distribution at higher order. Let us now turn to the discussion about the final state soft photon radiation effect. Since the emitted soft photon tends to be aligned with the outgoing leptons, the total transverse momentum of the lepton pair acquired from the recoil effect therefore also points toward the individual lepton's direction, on average.  This naturally generates positive $\cos\,2\phi$ and $\cos \, 4\phi$ asymmetries of purely perturbative origin.  The corresponding physics from such final state photon radiation is captured by the soft factor that enters the cross section formula via, 
  \begin{eqnarray}
    \frac{d \sigma(q_\perp)}{d {\cal P.S.}}=\int d^2 l_\perp \frac{d \sigma_0(q_\perp-l_\perp)}{d {\cal P.S.}}
    S(l_\perp),
  \end{eqnarray}
where $\sigma_0 $ is the leading order Born cross section given in Eq.~\eqref{born} and $d {\cal P.S.}$ stands for the phase space factor.  

In the small lepton mass $m\ll M$ limit,
 the soft factor   at the leading order can be expanded~\cite{Hatta:2020bgy,Hatta:2021jcd}, 
\begin{eqnarray}\label{eq:softfactor_approx}
S(l_{ \perp})\!=\! \delta(l_{ \perp})\!+\! \frac{\alpha_e }{\pi^2 l_{ \perp}^2} \!\left \{ c_0\!+\!2 c_2 \cos 2\phi_l\!+\!2 c_4\cos 4\phi_l\!+... \right \} , \,\,
\label{inte}
\end{eqnarray}
where $\phi_l$ is the angle between $P_\perp$ and the soft photon transverse momentum $-l_{ \perp}$.   When $y_1=y_2$, one has $c_0\approx \ln \frac{M^2}{m^2}$, $c_2\approx \ln \frac{M^2}{m^2}-4 \ln 2$ and $c_4\approx \ln \frac{M^2}{m^2}-4 $.  

Following the standard procedure, the soft factor in Eq.~\eqref{inte} can be extended to all orders by exponentiating the azimuthal independent part to the Sudakov form factor in the transverse position space. The resummed cross section takes the form~\cite{Catani:2014qha,Catani:2017tuc,Hatta:2020bgy,Hatta:2021jcd},
\begin{widetext}
 \begin{eqnarray}\label{eq:resum}
  \frac{d\sigma(q_\perp)}{d {\cal P.S.} }\!&=&\!\! \!\int \!
  \frac{d^2 r_\perp}{(2\pi)^2}
  \left [1\!-\!\frac{2\alpha_e c_2 }{\pi} \cos 2\phi_r +\frac{\alpha_e c_4}{\pi} \cos 4\phi_r\right ]  e^{i r_\perp \cdot q_\perp} e^{- \mathrm{Sud}(r_\perp)} \!\! \int \!\! d^2 q_\perp'
  e^{i r_\perp \cdot q_\perp'}  \frac{d\sigma_0(q_\perp')}{d {\cal P.S.} }.~~~
  \label{massless}
  \end{eqnarray}
\end{widetext}
Here $\phi_r$ is the angle between $r_\perp$ and $P_\perp$. The Sudakov factor at one loop is given by~\cite{Hatta:2020bgy,Hatta:2021jcd},
  \begin{eqnarray}\label{eq:Sudakov_exp}
  \mathrm{Sud}(r_\perp)=
\frac{\alpha_e}{\pi} {\rm ln} \frac{M^2}{m^2}  {\rm ln}\frac{P_\perp^2}{\mu_r^2}   
  \end{eqnarray}
with $\mu_r=2 e^{-\gamma_E}/|r_\perp|$.  We can  use Eq.~\eqref{massless} to compute the azimuthal asymmetries in di-muon production at LHC since the contributions suppressed by the power of $m^2/M^2$ can be safely neglected. This resummation formalism has been applied to study the lepton-jet correlation at the EIC~\cite{Tong:2022zwp} as well.

\begin{figure}[hbt]\centering
    \includegraphics[scale=0.3]{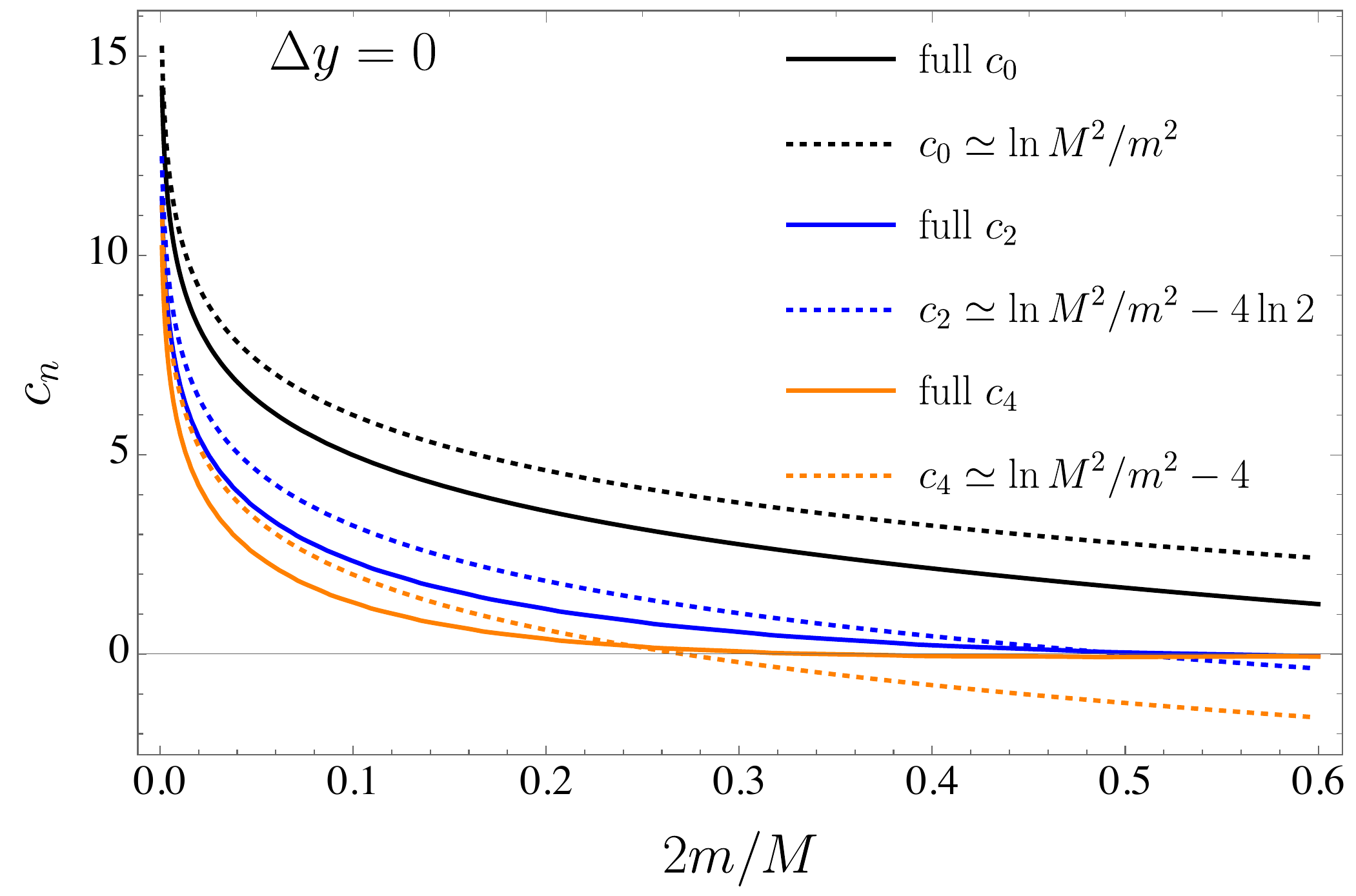}
    \caption{The coefficients of $c_0$ (black lines), $c_2$ (blue lines) and $c_4$ (red lines) in the soft factor, where the dashed lines are approximated results in Eq. \eqref{eq:softfactor_approx}, and the solid lines are obtained from the full expression in Eq. \eqref{eq:Sudakov_full}. Here we choose $\Delta y =0$. }
    \label{fig:c2c4}
\end{figure}  

However, at RHIC energy, where the lepton mass $m$ is roughly the same order of $M$, the soft factor receives the sizable finite lepton mass correction. In the soft photon limit $E_\gamma \ll m\sim M$, we consider the outgoing lepton with momentum $p_i^\mu = m v_i^\mu$. By taking the eikonal approximation, for each attachment of a photon to a lepton with velocity $v_i$, we have a factor $v_i^\mu/v_i\cdot k$. By contracting the pair of eikonal factors with the cut photon propagator and the transverse momentum measurement function, we have the one-loop soft factor in the Feynman gauge as
\begin{align}
  S(l_\perp,m,M)=& \sum_{i,j}\int \frac{d^d k}{(2\pi)^{d-1}} \delta(k^2)\theta(k^0) \frac{e^2 v_i\cdot v_j \Pi_{ij}}{v_i \cdot k \, v_j\cdot k}\notag \\
  & \times \delta^{(2)}(l_\perp-k_\perp),
\end{align}
where $i,j=1,2$, and the sign factor $\Pi$ is given by $\Pi_{ij}=+1$ if, $i\neq j$, while $\Pi_{ij}=-1$, if $i= j$. It is obviously symmetric in the indices $i$ and $j$.  This $d$-dimensional momentum integral can be evaluated in the transverse position space. Explicitly, one has the resumed formula as
\begin{align}\label{eq:Sudakov_full}
  e^{-\mathrm{Sud}(r_\perp)}\left[1+ \frac{\alpha_e}{4\pi}\left(s_{11} + s_{22} + 2 s_{12}\right) \right],
\end{align}
where the one-loop Sudakov factor is given by
\begin{align}
    \mathrm{Sud}(r_\perp) = \frac{\alpha_e}{\pi}\ln \frac{P_\perp^2}{\mu_r^2}\left(-1-\frac{1+\beta^2}{2\beta}\ln\frac{1-\beta}{1+\beta}\right),
\end{align}
and $s_{ij}$ are related to products of different eikonal factors, and they are given by
\begin{align}
  s_{11}&= s_{22}  = \frac{4 c_{r}}{\sqrt{c_{r}^{2}+1}} \ln \left(\sqrt{c_{r}^{2}+1}+c_{r}\right), \\
  s_{12}  & = -\frac{1+\beta^{2}}{2\beta} \operatorname{sign}\left(c_{r}\right)\Big[L_{\zeta}\left[\zeta\left(c_{r}, \alpha_r\right), \alpha_r\right] \notag \\
  & -L_{\zeta}\left[\zeta\left(-c_{r}, \alpha_r\right), \alpha_r\right]\Big],
\end{align}
with
\begin{align}
& c_{r}=\cos \phi_{r} P_\perp/m, ~~\beta=\sqrt{1-4m^2/M^2}   , \notag \\
  &\alpha_r=  \frac{2P_\perp^2 \cos ^{2} \phi_{r}}{-m^2+P_\perp^2 + (m^2+P_\perp^2)\cosh(y_1-y_2)},\notag \\
  &\zeta(a, b)=\left(a+\sqrt{1+a^{2}}\right)\left(a+\sqrt{a^{2}+b}\right), \notag\\ 
  &L_{\zeta}(a, b)=2 \bigg[-\operatorname{Li}_{2}\left(\frac{a+b}{b-1}\right)+\operatorname{Li}_{2}(-a) \notag \\
  & +\ln (a+b) \ln (1-b)\bigg]-\ln ^{2}\left(\frac{a}{a+b}\right)+\frac{1}{2} \ln ^{2}\left[\frac{a(a+1)}{a+b}\right]. \notag
\end{align}
To obtain the above finite results, we have applied the $\overline{\rm MS}$ subtraction scheme to remove all UV poles in the dimensional regularization and choose the renormalization scale as $P_\perp$. We note that the above soft integral was first calculated in \cite{Zhu:2012ts,Li:2013mia,Catani:2014qha} when the authors studied the transverse momentum resummation for heavy quark pairs production at hadron colliders. Besides, we also check that Eq. \eqref{eq:softfactor_approx} has included all logarithmic terms of $\ln\, m^2/M^2$. E.g., in the small lepton mass $m\ll M$ limit, the soft integral $s_{12}$ given in \eqref{eq:Sudakov_full} does indeed reduce to the expression in Eq. \eqref{eq:softfactor_approx}. In Fig.\ref{fig:c2c4}, we present the numerical results for the coefficient $c_2$ and $c_4$, where the dashed lines are obtained from the approximated expression in Eq. \eqref{eq:softfactor_approx}, and the solid lines are the full results in Eq. \eqref{eq:Sudakov_full}. We can see that in the small $m\ll M$ limit, the solid and dashed lines agree with each other, and as the increase of mass ratio, the power corrections become more and more important. Therefore, at the RHIC energy, we will apply the new resummation formula, including power corrections of $m^2/M^2$ at the one-loop order.

\section{Numerical results}
The azimuthal asymmetries, i.e., the average value of  $\cos (n\phi)$ that we are going to estimate numerically are defined as,
\begin{eqnarray}
\langle \cos(n\phi) \rangle &=&\frac{ \int \frac{d \sigma}{d {\cal P.S.}} \cos (n\phi) \ d {\cal P.S.} }
{\int \frac{d \sigma}{d {\cal P.S.}}  d {\cal P.S.}}
\end{eqnarray}
We compute the asymmetries for 60-80\% centrality region as well as for the unrestricted UPC events.
 The corresponding  impact parameter range for a given centrality class is determined by  using the Glauber model~\cite{Miller:2007ri}. For the UPC case,  we simply carry out $b_\perp$ integration  over the range $[2R_{\rm WS}, \infty)$, with the nucleus radius $R_{\rm WS}$ being 6.4 fm for Au and 6.68 fm for Pb. For the Au-Au 60-80\% centrality case, the $b_\perp$ integration range is [11.4 fm, 13.2 fm]. 

At the low transverse momentum, the photon distribution is dominated by the coherent ones that couple with the colliding nuclei as a whole. The coherent photon distribution is commonly computed with the equivalent photon approximation (also often referred to as  the Weizs$\ddot{a}$cker-Williams method), which has been widely used to compute UPC observables. In the equivalent photon approximation, $ {\cal F}(x,k_\perp)$ reads,   
  \begin{eqnarray}
    {\cal F}(x,k_\perp^2)=\frac{Z \sqrt{\alpha_e}}{\pi} |k_\perp|
   \frac{F(k_\perp^2+x^2M_p^2)}{(k_\perp^2+x^2M_p^2)},
  \end{eqnarray}
where $M_p$ is the proton mass. The nuclear charge density distribution in momentum space is taken from the STARlight generator,
\begin{eqnarray}
F(\vec k^2)=\frac{3[\sin(|\vec k|R_{A})-|\vec k|R_{A}\cos(|\vec k|R_{A})]}{(|\vec k|R_{A})^{3}(a^{2}\vec k^{2}+1)},
\end{eqnarray}
with $a=0.7$ fm and $R_{A}=1.1A^{1/3}$ fm. Such a parametrization is very close to the Fourier transform of the Woods-Saxon distribution numerically.

\begin{widetext}

\begin{figure}[hbt]\centering
    \includegraphics[scale=0.33]{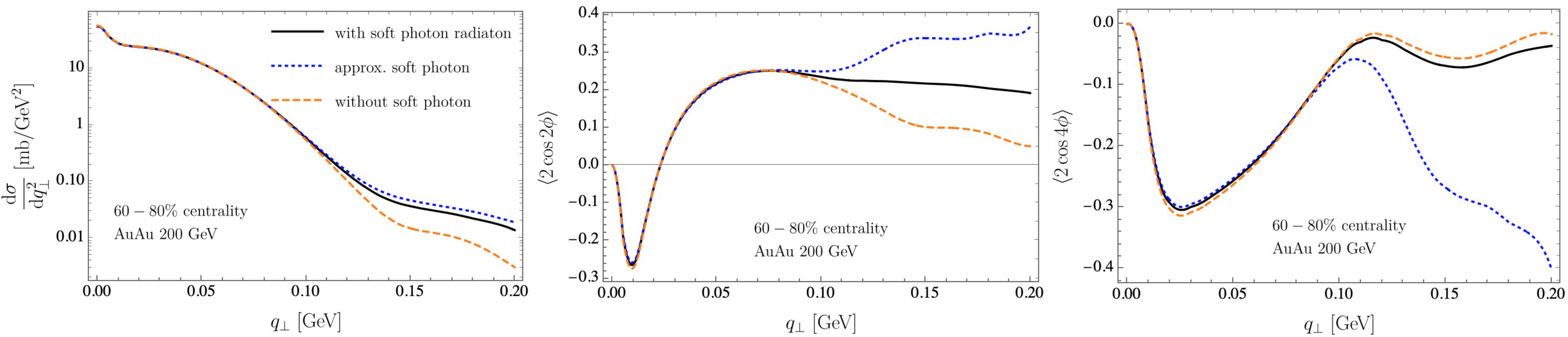}
    \caption{Di-muon production for $60 - 80\%$ centrality  in Au-Au collisions at the RHIC. The following kinematic cuts are imposed: the muons' rapidities $|y_{1,2}|<0.8$, transverse momentum $P_\perp >200$ MeV, and the invariant mass of the di-muon  $400~\text{MeV}< M < 640~\text{MeV}$.}
    \label{Au6080}
    \end{figure}     

    \begin{figure}[hbt]\centering
    \includegraphics[scale=0.33]{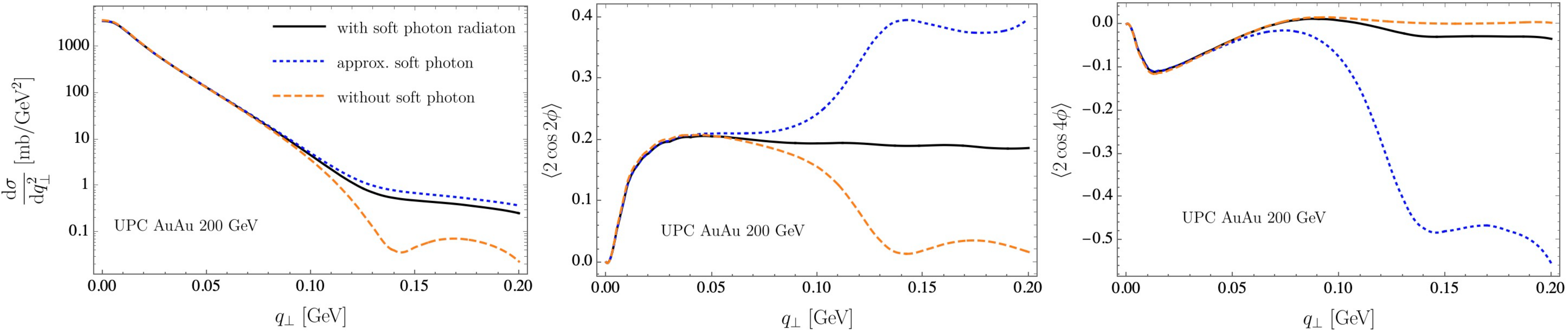}
    \caption{Di-muon production in unrestricted UPCs in Au-Au collisions at the RHIC.  The kinematic cuts are the same as given in Fig. \ref{Au6080}. 
    }
    \label{AuUPC}
    \end{figure}  

    \begin{figure}[hbt]\centering
    \includegraphics[scale=0.33]{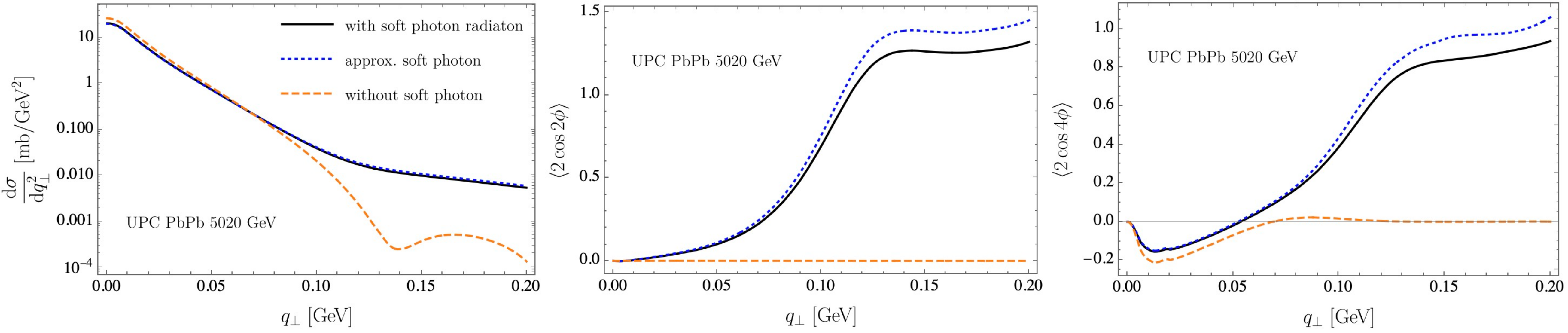}
    \caption{Di-muon production in unrestricted UPCs in Pb-Pb collisions at the LHC. The following kinematic cuts are imposed: the muons' rapidities $|y_{1,2}|<1$, transverse momentum $P_\perp >4$ GeV, and the invariant mass of the di-muon  $10~\text{GeV}< M < 45~\text{GeV}$.}
    \label{PbUPC}
    \end{figure} 
\end{widetext}

The numerical results for the unpolarized cross section of di-muon  production as well as the $\cos 2\phi$ and $\cos 4\phi$ azimuthal asymmetries for the $60\%-80\%$ centrality region at RHIC energy are
 presented in Fig. \ref{Au6080}.  Note that when computing the asymmetries with/without soft photon contributions, the denominator is always the complete resummed unpolarized cross section. The predictions  for the unrestricted UPC events at RHIC are shown  in Fig.~\ref{AuUPC}.  The asymmetries at low transverse momentum are mainly induced by the primordial linearly polarized photon distribution, while they are dominated by the final state soft photon radiation effect at relatively high pair transverse momentum ($q_\perp>100$ MeV).  One can clearly see that the contribution to the asymmetries from the muon mass effect incorporated in the resummation formalism  is rather sizable at high pair transverse momentum. To be more specific, at RHIC energy the mass correction effect tends to reduce $\cos 2\phi $ and $\cos 4\phi $ asymmetries at large $q_\perp$ as compared to the resummed results obtained without considering the mass effect. This is expected because the soft photon emissions are more mildly peaked around the outgoing charged particle direction for a massive emitter.  
It would be interesting to  test this theory predication against  the future measurement at the RHIC. The computed asymmetries at RHIC energy are also shown as the function of the invariant mass in Fig.~\ref{Au6080M} for 60\%-80\% centrality  and Fig.~\ref{AuUPCM} for the UPC case.

 We display the unpolarized cross section of di-muon  production at LHC together with the computed asymmetries in Fig.~\ref{PbUPC}. As we know, the $\cos\, 2\phi$ azimuthal modulation  arises from the linearly polarized photon distribution  is proportional to $m^2/P_\perp^2$~\cite{Li:2019yzy},  which is negligibly small at LHC energy.  Therefore the  $\cos \, 2\phi$ asymmetry is entirely  generated from the final state soft photon radiation effect.  One also observes that the muon mass effect entering the resummation formalism leads to very mild corrections to the asymmetries at LHC energy as it is suppressed by the power of  $m^2/M^2$.

\begin{widetext}
    
   \begin{figure}[hbt]\centering
\includegraphics[scale=0.33]{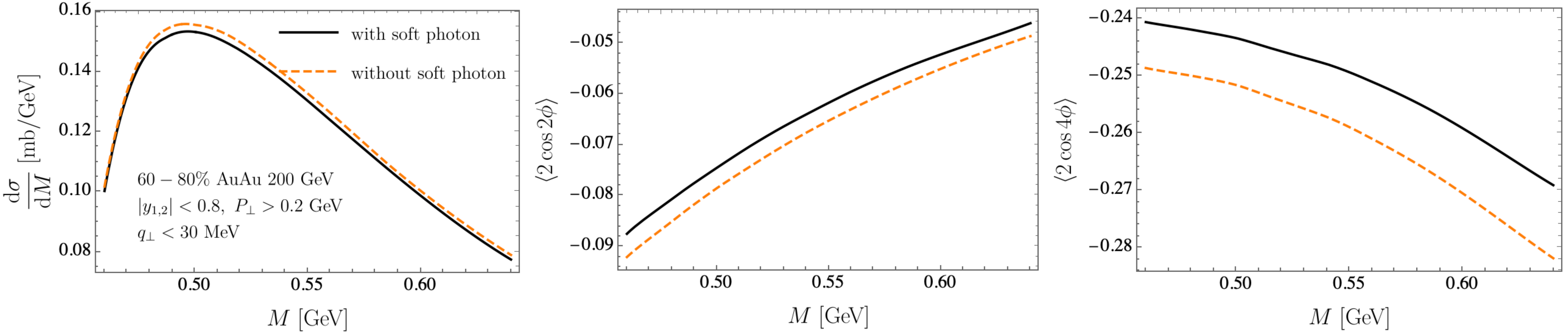}
    \caption{Di-muon production for 60\%-80\% centrality in Au-Au collisions at the RHIC, as the function of the invariant mass. 
    }
    \label{Au6080M}
    \end{figure}
 
   \begin{figure}[hbt]\centering
\includegraphics[scale=0.33]{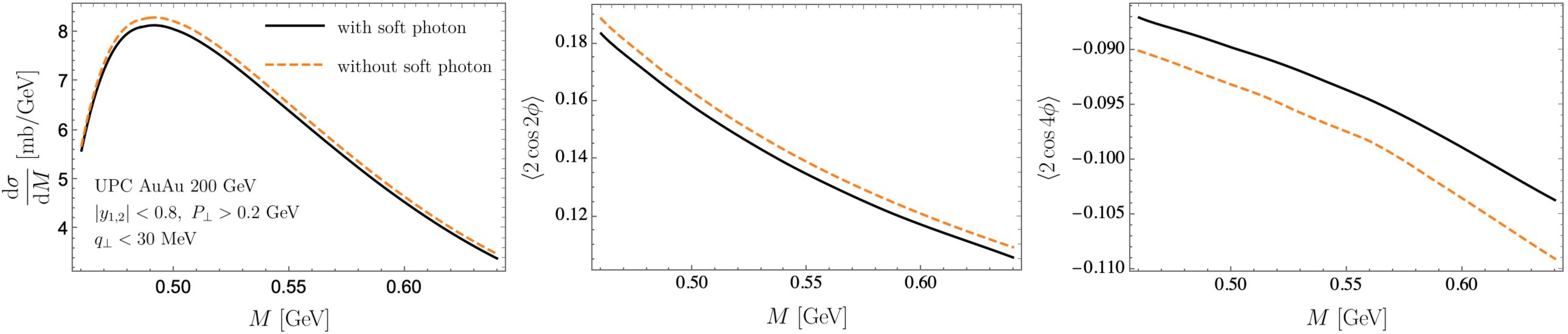}
    \caption{Di-muon production in unrestricted UPCs in Au-Au collisions at the RHIC, as the function of the invariant mass. 
    }
    \label{AuUPCM}
    \end{figure}
\end{widetext}
\ 

\section{Conclusion}\label{Conclusion}
In this work, we study the azimuthal asymmetries in di-muon production via the photon fusion process in UPCs.  At the low pair transverse momentum, the asymmetries are mainly induced by the linearly-polarized coherent photons. Compared to the previous calculation~\cite{Li:2019yzy}, we improved the analysis  by taking into account the impact parameter dependence.  The primordial coherent photon distribution decreases exponentially at large $q_\perp$, where the perturbative tail generated by the final state soft photon radiation dominates the distribution.  As soft photons are most likely emitted along the produced muon direction, they naturally lead to the sizable $\cos \, 2\phi$ and $\cos\, 4\phi$  azimuthal asymmetries as well.  Such soft photon radiations are resummed to  all orders following the  approach~\cite{Zhu:2012ts,Li:2013mia,Catani:2014qha} initially developed for computing heavy quark pair production in hadron collisions. Compared to the previous study~\cite{Hatta:2021jcd}, the resummation scheme employed in the current calculation allows us to take into account the full finite lepton mass correction. Though its correction to the unpolarized cross section is tiny, our numerical results indicate that the contribution from the muon mass effect to the asymmetries is quite sizable at large $q_\perp$ at RHIC energy. At LHC energy, such effect is negligible due to its power correction nature.  The azimuthal asymmetries in di-muon production at RHIC thus provide us a unique opportunity to test our understanding of the resummation technique beyond the leading power and the leading logarithm contributions in a very clean way.

\section*{Acknowledgments}
We thank Chi Yang, Xiao-feng Wang, Wang-mei Zha, Ze-bo Tang and Jian Zhou from USTC for helpful discussions. 
D.Y.S.~is supported by the National Science Foundations of China under Grant No.  12275052 and the Shanghai Natural Science Foundation under Grant No. 21ZR1406100. J. Zhou has been supported by the National  Science Foundations of China under Grant No.\ 12175118. Y. Zhou has been supported by the Natural  Science Foundation of Shandong Province under Grant No. ZR2020MA098. C. Zhang has been supported by the National Science Foundations of China under Grant No.\ 12147125.
 
\bibliography{ref}
\end{document}